\pdfoutput = 1

\documentclass[a4paper,12pt]{article}
\usepackage{hyperref}

\usepackage{graphicx}
\usepackage{caption}
\usepackage{cite}
\usepackage{xcolor}
\usepackage{subfig}
\usepackage{slashed}
\usepackage{float}
\usepackage{ulem}
   
\restylefloat{figure}

\usepackage[sumlimits,intlimits,namelimits]{amsmath} 

\usepackage[english]{babel}

\usepackage{graphicx}

\usepackage{geometry}
\numberwithin{equation}{section}
\newcommand{\gev}{{\rm GeV}}
\newcommand{\Ref}[1]{Ref.~\cite{#1}}

\begin{document}
\begin{titlepage}
\begin{flushright}
SI-HEP-2017-15, QFET-2017-13 \\[0.2cm]
MIT-CTP-4926, TUM-1095/17, NIOBE-2017-03 \\[0.2cm]
\end{flushright}

\vspace{1.2cm}
\begin{center}
{\Large\bf 
CP Violation in Multibody \boldmath $B$ \unboldmath Decays \\[2mm] from  QCD Factorization}
\end{center}

\vspace{0.5cm}
\begin{center}
{\sc Rebecca Klein${}^{\,a}$, Thomas Mannel$^{\,a}$, Javier Virto$^{\,b,c}$, K. Keri Vos$^{\,a}$ } \\[0.3cm]
{\sf ${}^a$Theoretische Physik 1, Naturwiss. techn. Fakult\"at, \\
Universit\"at Siegen, D-57068 Siegen, Germany}\\
\vspace{0.4cm}
{\sf ${}^b$Physik Department T31, Technische Universit\"at M\"unchen, \\
James-Franck-Stra{\ss}e 1, D-85748 Garching, Germany}\\
\vspace{0.4cm}
{\sf ${}^c$Center for Theoretical Physics, Massachusetts Institute of Technology, \\
77 Mass. Ave., Cambridge, MA 02139, USA}

\end{center}

\vspace{0.8cm}
\begin{abstract}
\vspace{0.2cm}\noindent

\noindent We {test} a data-driven {approach} based on QCD factorization for charmless three-body $B$-decays
by confronting it to measurements of CP violation in $B^- \to \pi^- \pi^+ \pi^-$.
While some of the needed non-perturbative objects can be directly extracted from data,
some others can, so far, only be modelled.  Although this approach is currently model dependent, we comment on the perspectives to reduce this model dependence. 
While our model naturally accommodates
the gross features of the Dalitz distribution, it cannot quantitatively explain the details seen in the 
current experimental data on local CP asymmetries. We comment on possible refinements of our simple model and conclude by briefly discussing a possible extension of the model to large invariant masses, where large local CP asymmetries have been measured.

\end{abstract}

\end{titlepage}

\newpage
\pagenumbering{arabic}
\section{Introduction}
Hadronic multi-body decays with more than two final-state hadrons constitute a large part of the branching fraction 
for heavy hadron non-leptonic decays.  In principle, three- and more body decays have non-trivial kinematics
and the phase space distributions contain far more information than the two-body decays. 

For $D$ decays there exists a large amount of data on multi-body decays. However, the charm quark mass is not large 
enough for heavy quark methods, since the typical invariant masses $m_{ij}$ of final state hadron pairs are roughly $m_c / \sqrt{N}$, where $N$ is the 
final state multiplicity. Already for three-body decays of charmed hadrons this is outside the perturbative region and hence there is no chance 
to discuss the resulting amplitudes on the basis of some factorization theorem. 

For $B$ decays the situation may be slightly better, as we pointed out in a recent publication~\cite{Kra15}. For three-body decays 
such as $B \to \pi \pi \pi$  the bottom-quark mass turns out to be still too small to allow for a complete factorization in the central region of the Dalitz plot. However, it seems 
that one can make use of a ``partial factorization'' at the edges of the Dalitz distribution, where the invariant mass of two of the pions is 
small. 

It has been discussed in \cite{Kra15,BenekeTalk} that for this part of the phase space the same proof of factorization as for the 
two-body decays \cite{BBNS, BBNS2, Ben01, BenNeu03} is valid. However, the non-perturbative input given by the matrix elements of the factorized operators 
is different: In the case of three-body decays the light-cone distribution of two collinearly moving pions and the soft $B \to \pi \pi$ form factor
are needed.  At least the edges of the Dalitz plot can be described in terms of these quantities; however, as has been argued 
in \cite{Kra15} this formalism {may} be extrapolated to the more central parts of the Dalitz plot.  
   
Hadronic multi-body decays are also interesting for studies of CP violation. Although the integrated (direct)  CP asymmetries are 
small, local CP asymmetries (i.e. the CP asymmetry for fixed values of the final-state invariant masses) are measured to be large 
in some regions of the phase space and exhibit a rich structure \cite{LHCbdata, Lee15, Aub09, Aai13, Aai14, Gar05}.  Assuming the 
well known CKM mechanism for CP violation, the requirement for its appearance is an interference between at least two amplitudes 
with different weak and strong phases. Since the weak phases are independent of the kinematics, any dependence on the kinematical 
variables of the local CP asymmetries reflects kinematics-dependent strong-phase differences. 

In the present paper we discuss an approach for three-body decay amplitudes based on QCD factorization. 
We will take into account the leading term only, which  is equivalent to adopting naive factorization for the hadronic matrix elements. 
The main purpose is to study to which extent such a framework can properly describe the observed Dalitz distribution and local CP asymmetries in $B^- \to \pi^- \pi^+ \pi^-$.

In the next section we  summarize the QCDF formula for three-body decays, then we discuss the  non-perturbative input
needed in the factorization formula.   
In section 4 we compute the Dalitz distributions and the local CP asymmetries in our framework, with a fit to experimental data.  
We conclude with a discussion of the results.

\section{QCD-Factorization for $B^-  \to \pi^-  \pi^+  \pi^-$}
In the following we will discuss charmless hadronic three-body decays and as a concrete example we consider $B^-  \to \pi^-  \pi^+  \pi^-$. We define the external momenta
\begin{equation}
B^-(p_B)  \to \pi^-(k_1) +  \pi^+(k_2) + \pi^-(k_3) \ ,
\end{equation}
where $p_B=k_1 + k_2 + k_3$ and, for massless pions,
\begin{equation}
p_B^2 = m_B^2, \quad k_i^2 = 0, \quad s_{ij} \equiv \frac{(k_i+k_j)^2}{m_B^2} \ ,
\end{equation}
such that $s_{12}+s_{13}+s_{23}=1$. For $B^- \to \pi^-\pi^+\pi^-$ the Dalitz distribution is symmetric in $s_{12}$ and $s_{23}$. Experimentally these variables cannot be distinguished, and we {define $k_1$ and $k_3$ by} $s_\pm^{\rm low}\equiv s_{12}$ and $s_\pm^{\rm high}\equiv s_{23}$,
with $s_\pm^{\rm low} < s_\pm^{\rm high}$.

The application to other combinations of charges as well as to final states with kaons is obvious.   
As we have discussed in our previous paper~\cite{Kra15}, the structure of the
amplitude in the region $s_\pm^{\rm low}\ll 1$ is very similar to the two body case within the
QCD-factorization framework.
The only difference is that the matrix elements of the operators will eventually induce new non-perturbative quantities. 

In this region, the $B^- \to \pi^-\pi^+\pi^-$ amplitude at leading order in $\alpha_s$ and at leading twist is given by~\cite{Kra15}
\begin{align}\label{eq:ampli}
\mathcal{A}(s_{\pm}^{\rm low} , s_{\pm}^{\rm high}) {}& = \frac{G_F}{\sqrt{2}}
\left\{
\big[ \lambda_u(a_2 - a_4^u) -\lambda_c a_4^c \big] \, m_B^2\, f_+(s_{\pm}^{\rm low} )\,(1-s_{\pm}^{\rm low} -2 s_{\pm}^{\rm high}) \,F^{\textrm{em}}_\pi(s_{\pm}^{\rm low} ) \right.\nonumber \\
{}& \left. + \big[\lambda_u(a_1 +a_4^u ) + \lambda_c a_4^c \big] \,f_\pi\, m_\pi\,  \big[F_t^{I=0}(s_{\pm}^{\rm low},s_{\pm}^{\rm high})+F_t^{I=1}(s_{\pm}^{\rm low},s_{\pm}^{\rm high}) \big] 
\right\} \ .
\end{align}
The quantities $\lambda_p\equiv V_{pb} V_{pd}^*$ encode the CKM factors,
$a_{1,2}$ and $a_4^{u,c}$ are constructed from Wilson coefficients, loop functions and convolutions with light-cone distributions (see Sec.~\ref{sec:CPV} and \Ref{BenNeu03}),
and the objects $f_\pi, f_+, F^{\rm em}_\pi$ and $F_t^{I}$ are non-perturbative quantities to be discussed 
in Sec.~\ref{sec:NPinput}.
The amplitude in Eq.~\eqref{eq:ampli} is the key formula in this paper.

\bigskip

At this order and twist, this formula coincides with the result obtained by applying the ``naive-factorization" ansatz (see e.g. \cite{Ded11}), and we find it convenient to use some of its notation here.
To this end, we can simply take the QCD-factorized effective Hamiltonian from \cite{BBNS}, which reads:
\begin{equation}
 {\cal H}_{\textrm{eff}}  = \frac{G_F}{\sqrt{2}} 
\left( \lambda_u T_u + \lambda_c T_c \right) 
\end{equation}
with 
\begin{eqnarray}
T_u  &=&  a_1^u   \,\,  \left[  (\bar{u}b)_{V-A} \times (\bar{d}u)_{V-A} \right]  \vphantom{\sum_q} 
              + a_2^u \,\, \left[ (\bar{d}b)_{V-A}  \times (\bar{u}u)_{V-A}  \right] \vphantom{\sum_q}  \nonumber \\
              &+& a_3  \,\, \sum_q \left[ (\bar{d}b)_{V-A} \times (\bar{q}q)_{V-A} \right] 
                + a_4^u \,\, \sum_q  \left[ (\bar{q}b)_{V-A}  \times (\bar{d}q)_{V-A} \right]
\nonumber \\
               &+& a_5 \,\, \sum_q  \left[ (\bar{d}b)_{V-A} \times (\bar{q}q)_{V+A} \right] 
                - 2  a_6^u \,\, \sum_q \left[ (\bar{q}b)_{S-P} \times (\bar{d}q)_{S+P} \right] \, , 
\end{eqnarray} 

\begin{eqnarray}
T_c  &=&  a_3  \,\, \sum_q \left[ (\bar{d}b)_{V-A} \times (\bar{q}q)_{V-A} \right] 
              + a_4^c \,\, \sum_q \left[ (\bar{q}b)_{V-A}  \times (\bar{d}q)_{V-A} \right] \nonumber \\
             &+& a_5 \,\, \sum_q \left[ (\bar{d}b)_{V-A} \times (\bar{q}q)_{V+A} \right]
                   - 2  a_6^c \,\, \sum_q \left[ (\bar{q}b)_{S-P} \times (\bar{d}q)_{S+P} \right] \ . 
\end{eqnarray}
The notation of the operators means that the matrix element is to be evaluated in the 
factorized form as a product of two matrix elements. 
To be specific, for the case at hand we have the two cases
\begin{eqnarray}
&&\hspace{-2cm} \langle \pi^- (k_1)  \pi^+ (k_2)  \pi^- (k_3) |  \left[  (\bar{u}b)_{V-A} \times (\bar{d}u)_{V-A} \right] | B^- (p_B) \rangle 
 \nonumber \\[1mm]
&& =  \langle \pi^- (k_1)  \pi^+ (k_2) | (\bar{u}b)_{V-A}   | B^-(p_B) \rangle  \,  \langle \pi^- (k_3)  | (\bar{d}u)_{V-A} | 0 \rangle +  
k_1 \leftrightarrow k_3  \vphantom{\sum_q}\ ,
\label{fact1}\\
&&\hspace{-2cm} \langle  \pi^- (k_1)  \pi^+ (k_2) \pi^-(k_3)| \left[ (\bar{d}b)_{V-A}  \times (\bar{u}u)_{V-A}  \right]  | B^- (p_B) \rangle 
\nonumber \\[1mm]  
&&= \langle \pi^- (k_3)  |  (\bar{d}b)_{V-A}  | B^- (p_B) \rangle \, \langle \pi^- (k_1) \pi^+ (k_2) |  (\bar{u}u)_{V-A} | 0 \rangle 
+ k_1 \leftrightarrow k_3 \vphantom{\sum_q}\ .
\label{fact2} 
\end{eqnarray}

The relevant non-perturbative objects in the leading-order amplitude are 
the pion decay constant {$f_\pi$},  the $B \to \pi$ form factor {$f_+$}, {the time-like helicity} $B \to \pi \pi$ form factors {$F_t^{I=0}$ and $F_t^{I=1}$}, and the 
pion form factor in the time-like region $F_\pi$. In the following section we give proper definitions
for these objects and specify how they will be fixed in our approach.

\section{Non-perturbative Input} 
\label{sec:NPinput}
The strength of our QCD-factorization based model is that non-perturbative inputs may be obtained from data. The pion decay constant and the $B \to \pi$ form factors can both be taken as real, but the new $B\to\pi\pi$ and pion form factors contain non-perturbative strong phases which will be driving the CP asymmetry distribution. 

\subsection{The pion decay constant and the timelike pion form factors} 
We define the pion decay constant in the usual way  
\begin{equation} \label{Pidec}
  \langle \pi^- (k_3)   | (\bar{d}u)_{V-A} | 0 \rangle = - \langle \pi^- (k_3)  | \bar{d} \gamma^\mu \gamma_5 u  | 0 \rangle = i f_\pi k_3^{ \mu}  \ ,
\end{equation}
with the numerical value $f_\pi \sim 130$ MeV. 

\begin{figure}
  \centering
  \subfloat[]{
\includegraphics[width=0.48\textwidth,height=6.5cm]{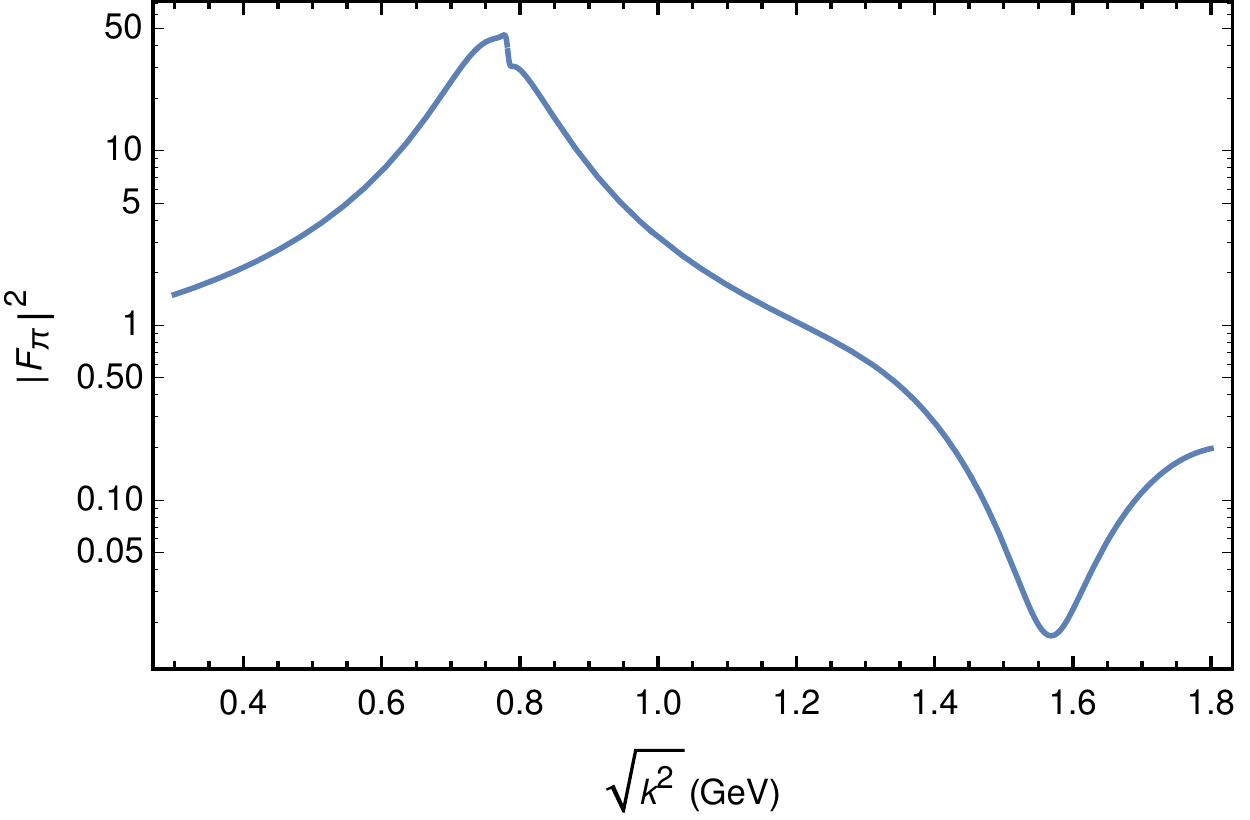}}
\hspace{3mm}
    \subfloat[]{\includegraphics[width=0.47\textwidth,height=6.5cm]{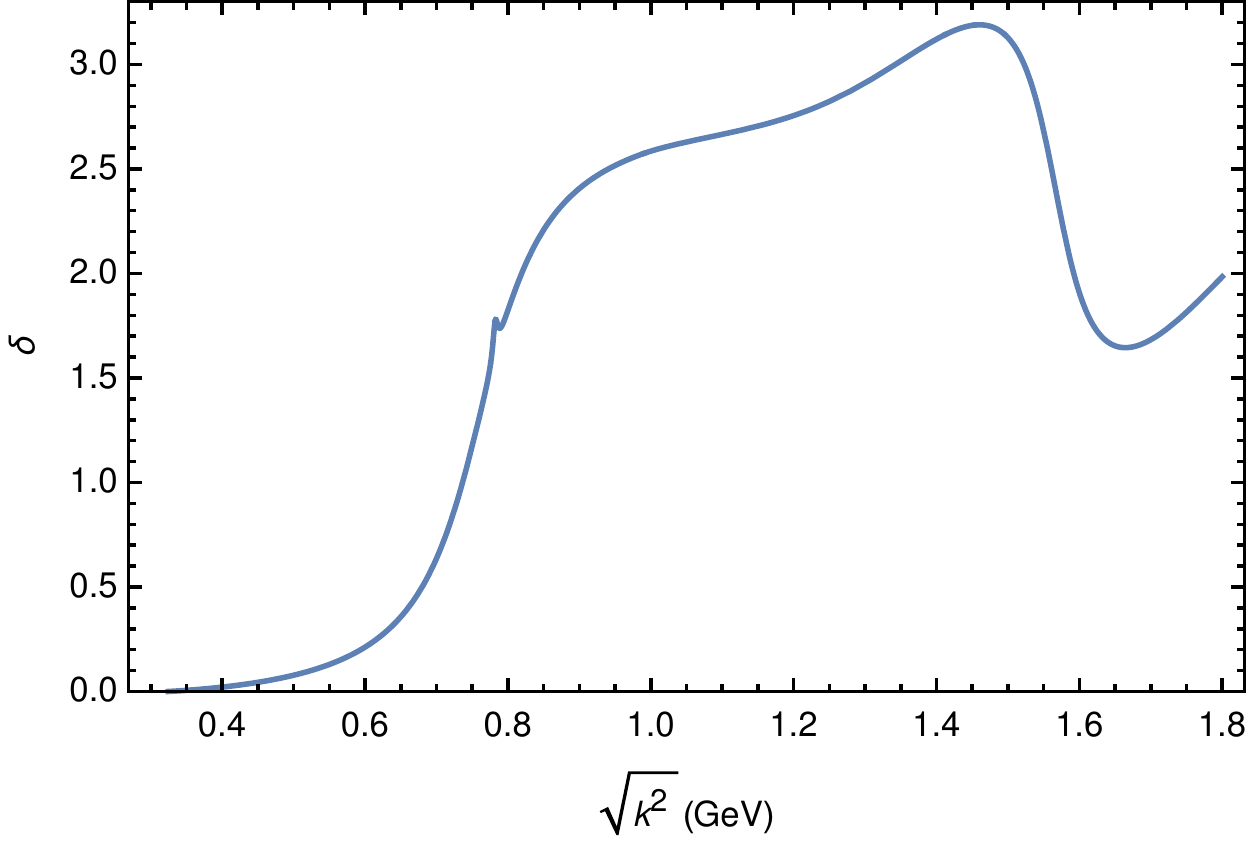}}
  \caption{Pion vector form factor $F_\pi^{\textrm{em}}(k^2)=|F_\pi^{\textrm{em}}| e^{i \delta}$ in the time-like region.  }
  \label{fig:vectorff}
\end{figure}

The pion form factor is defined by 
\begin{equation} \label{eq:2pionconst}
\left\langle \pi^- (k_1)\pi^+ (k_2) | \bar{q} \gamma^\mu  q | 0 \right\rangle\ = F_\pi^{\rm em} (k^2) (k_1 - k_2)^\mu 
\quad , \quad k^2 \equiv (k_1 + k_2)^2 \ge 0  
\end{equation} 
and can be obtained from electromagnetic probes.
Note that in the time-like region this form factor picks up a non-trivial  strong phase. Here we use the  parametrization of \Ref{She12}  {fitted to the} measurements of  $e^+e^- \to \pi^+\pi^-(\gamma)$ \cite{BaBar} (see also \Ref{Han12}). {The absolute value and the phase of this form factor are shown in Fig.~\ref{fig:vectorff}.} Unfortunately, {while the absolute value is very precisely measured up to $k^2\sim 3.5~\gev^2$, its phase is not so well constrained. This will add to the level of model dependence of our approach.}  

\begin{figure}
  \centering
  \subfloat[]{
\includegraphics[width=0.48\textwidth,height=6.5cm]{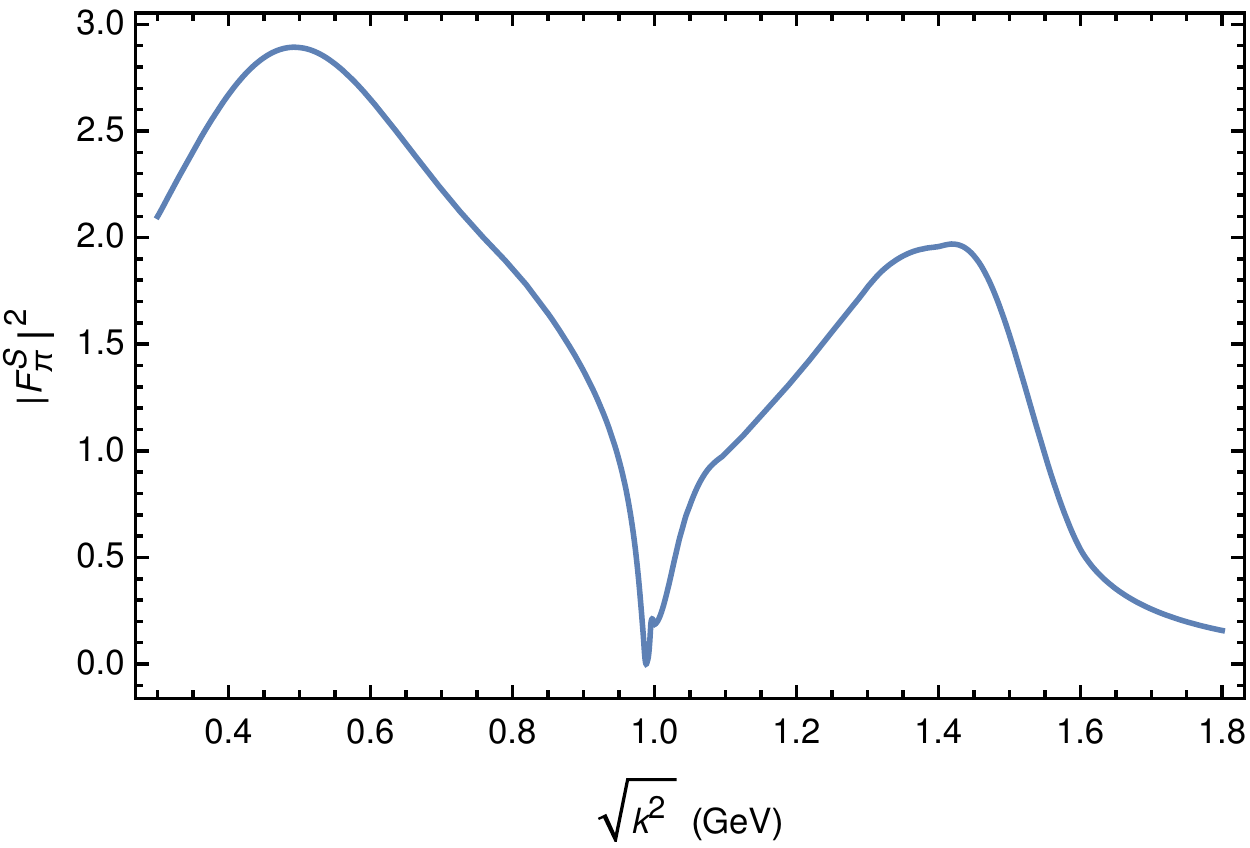}}
\hspace{3mm}
    \subfloat[]{\includegraphics[width=0.47\textwidth,height=6.5cm]{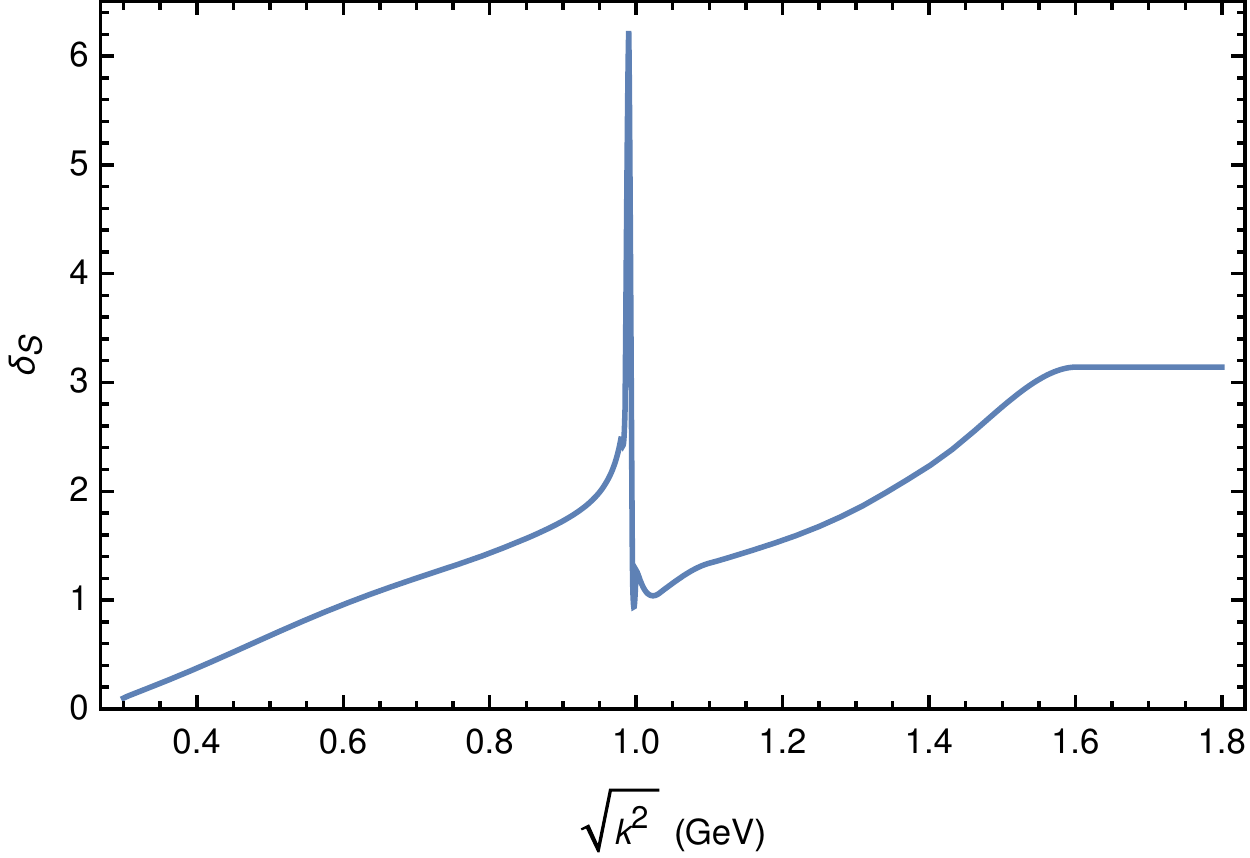}}
  \caption{Pion scalar form factor $F_\pi^S(k^2)=|F_\pi^S| e^{i \delta_S}$ in the time-like region.
  }
  \label{fig:ff}
\end{figure}

 {W}e will also need the corresponding form factor for the scalar current:  
\begin{equation}\label{eq:Fscalar}
\left\langle \pi^-(k_1)\pi^+(k_2) | m_u \bar{u}  u + m_d \bar{d}d | 0 \right\rangle = m_\pi^2 F_\pi^S(k^2) \ ,
\end{equation}
where the mass factors are chosen such that a proper chiral limit exists{~\cite{Daub15}}. This form factor can be obtained using a coupled channel analysis. We use the results of \Ref{Pas14}, which are valid up to around {$k^2 \simeq 3\,\gev^2$}, {as shown} in Fig.~\ref{fig:ff}. Similar results {have been} obtained in \Ref{Daub15} {in connection with a} study {on} $B \to J/\psi \pi\pi$. We note that the {shape} of $F_\pi^S(k^2)$ {around low-lying scalar resonances such as the $f_0(500)$} does not even remotely resemble the shape of a Breit-Wigner function. 

\subsection{The $B\to \pi$ form factor}
{We use the following definitions for the vector form factors~\cite{BenFel01}:}
\begin{equation}
\left\langle \pi^-(k_3)|\bar{d}\gamma^\mu b |B^-(p_B)\right\rangle\ = 
f_+(k^2) \left[ p_B^\mu + k_3^\mu - \frac{m_B^2-m_\pi^2}{k^2}k^\mu \right] + f_0(k^2) \frac{m_B^2 - m_\pi^2}{k^2}k^\mu  \ ,
\end{equation}
where $k  = p_B-k_3 =  k_1 + k_2 $. 

When applying the factorization formula (\ref{fact2}), this expression is contracted with the time-like 
vector form factor for the two other pions. Using the fact that the current of these two pions is conserved,  we get for (\ref{fact2})
\begin{equation}
\langle \pi^- (k_3)  |  (\bar{d}b)_{V-A}  | B^- (p_B) \rangle \, \langle \pi^- (k_1) \pi^+ (k_2) |  (\bar{u}u)_{V-A} | 0 \rangle  
= f_+ (k^2) F_\pi^{\rm em} ( k^2) \, \, 2k_3\cdot \bar{k} \ ,
\end{equation} 
where $\bar{k} \equiv k_1-k_2$. 
For the form factor $f_+$ we use the LCSR calculation in \Ref{Ims14}.

\subsection{The $B\to \pi\pi$ form factors}
The form factors appearing in the $B \to \pi \pi$ transitions have been studied in \cite{Fal14, Boe16} for $B\to \pi\pi \ell \nu$ and we use the 
definitions from these papers. However, when applying (\ref{fact1})  we only need the contraction with the 
matrix element (\ref{Pidec}), and hence only a single form factor appears
\begin{equation} \label{eq:Ftdef}
 \left\langle \pi^-(k_1)\pi^+(k_2)|\bar{u}\slashed{k}_3 \gamma_5 b|B^-(p_B)\right\rangle\  = i m_\pi F_t (k^2, k_3\cdot \bar{k})  \ ,
\end{equation}
where we used that $k_3^2=  m_\pi^2$, and
\begin{equation}\label{eq:cos}
k_3 \cdot \bar{k} = \frac{\beta_\pi}{2} \sqrt{\lambda} \cos\theta_\pi =  \frac{m_B^2}{2} (1-s_{\pm}^{\rm low} -2 s_{\pm}^{\rm high})
\end{equation}
defines the polar angle $\theta_\pi$ of the $\pi^-$ in the rest frame of the {dipion},
where $\beta_\pi^2 = (k^2- {4}m_\pi^2)/k^2$ and
$\lambda=\lambda(m_B^2, m_\pi^2, k^2) {= (m_B^2 - m_\pi^2 - k^2)^2 - 4 m_\pi^2 k^2}$ is the K\" all\' en function.  

The two pions can have isospin $I = 0$ or $I = 1$, such that
\begin{equation}
F_t = F_t^{I=0} + F_t^{I=1} \ .
\end{equation}

The {isovector form factor} $F_t^{I=1}$ {has been} studied using QCD {light-cone} sum rules in \cite{Che17,Che172},
{(see also \cite{Ham15} for a similar study of the other $P$-wave form factors).}
{Analogous studies of the isoscalar form factor $F_t^{I=0}$ have} not been performed.
Here we model the form factor  $F_t^{I=1}$ by assuming that the decay $B \to \pi \pi$ proceeds only {resonantly} through $B \to \rho \to \pi \pi$.  In this approximation
we need the form factors for the $B \to \rho$ transition via the left-handed current. In general, this requires four form 
factors, but when applying (\ref{fact1}) this reduces to a single form factor for the axial vector current
\begin{equation}
\left\langle \rho^0(k,\epsilon) | \bar{u}\gamma_\nu \gamma_5 b |B^-(p_B) \right\rangle 
= \frac{i}{\sqrt{2}} q_{\nu} (\epsilon^* \cdot q) \frac{2m_\rho}{q^2} A_0 (q^2) +\cdots \ , 
\end{equation}
where $\epsilon$ is the polarization vector of the $\rho$ meson with momentum $k$, and $q$ is the momentum transfer. 

Treating the $\rho$ as an intermediate resonance we obtain
\begin{eqnarray} 
&&\hspace{-2cm} \left\langle \pi^-(k_1)\pi^+(k_2)|\bar{u} \gamma_\nu (1-\gamma_5) b|B^-(p_B)\right\rangle =   \nonumber \\[1mm]
&& \quad \sum_\epsilon  \left\langle \pi^-(k_1)\pi^+(k_2) |\rho^0 (k,\epsilon)  \right\rangle {\cal B}_\rho (k^2) 
\left\langle \rho^0(l,\epsilon) | \bar{u}\gamma_\nu (1-\gamma_5) b |B^-(p_B) \right\rangle \ ,
\end{eqnarray} 
where we sum over the $\rho$ polarizations 
$$
 \sum_\epsilon \epsilon_\mu \epsilon_\nu^* =  - g_{\mu \nu} + \frac{k_\mu k_\nu}{k^2}  
$$
and introduce the Breit-Wigner function 
\begin{equation}\label{eq:BW}
{\cal B}_P (k^2) =\frac{1}{k^2 - m_P^2  + i \sqrt{k^2} \Gamma_P} \ ,
\end{equation}
where $\Gamma_P$ is the total decay width of the particle $P$. 

The decay matrix element for the $\rho \to \pi \pi$ transition is defined as 
\begin{equation} 
\left\langle \pi^-(k_1)\pi^+(k_2) |\rho^0\right\rangle = g_{\rho\pi^-\pi^+} (k_1-k_2)^\mu \epsilon_\mu \ ,
	\end{equation}
	and $g_{\rho\pi^-\pi^+}$ can be obtained from the decay width of the $\rho$ resonance. 
 
Combining {the various ingredients} and contracting with $q = k_3$ gives
\begin{equation} \label{BpipiVBW} 
\left\langle \pi^-(k_1)\pi^+(k_2)|\bar{u} \slashed{k}_3 (1-\gamma_5) b|B^-(p_B)\right\rangle 
 =\frac{2i m_\rho}{\sqrt{2}} g_{\rho\pi\pi}  (\bar{k}\cdot k_3)  A_0(m_\pi^2) \mathcal{B}_\rho(k)\ .
\end{equation}

Replacing the outgoing two pion state by a $\rho$ resonance described by a simple Breit-Wigner shape is clearly 
a crude approximation for both the absolute value {and} the phase.
{We refine this approximation in the following way: We use the same model for the
time-like form factor, and we determine a replacement for the Breit-Wigner function in terms of the measured pion form factor in Fig.~\ref{fig:vectorff}.}
{First, we have}:
\begin{align}
 \left\langle \pi^-(k_1)\pi^+(k_2)|\bar{u} \gamma_\nu(1-\gamma_5) u|0 \right\rangle{} &= (k_1-k_2)_\nu F^{\rm{em}}_\pi(k^2)  \\[2mm]
{} \quad &= \sum_\epsilon \left\langle \pi^-(k_1)\pi^+(k_2) |\rho^0 (k,\epsilon) \right\rangle {\cal B}_\rho (k) 
\left\langle \rho^0 (k,\epsilon)  |\bar{u}\gamma_\nu(1-\gamma_5)u|0 \right\rangle\ .
\nonumber
\end{align}
The {$\rho$-meson} decay constant is {then} defined by
\begin{equation}
\left\langle \rho^0|\bar{u}\gamma_\mu(1-\gamma_5)u|0\right\rangle = \frac{1}{\sqrt{2}}f_\rho m_\rho \epsilon^*_\mu 
\end{equation}
which allows us to write the pion form factor as 
\begin{equation}\label{eq:Fpirhomod}
F_\pi^{\textrm{em}}(k^2) = \frac{- f_\rho m_\rho \;g_{\rho \pi^+\pi^-} }{\sqrt{2}}\mathcal{B}_\rho(k^2)  \ .
\end{equation} 
We can now solve for $g_{\rho \pi \pi} {\cal B}_P$ 
and insert this into (\ref{BpipiVBW}). Finally, using (\ref{eq:Ftdef}) yields
\begin{equation}\label{eq:Ft1mod}
  F_t^{I=1}(k^2, k_3\cdot\bar{k}) = 2k_3\cdot \bar{k} \frac{F_\pi^{\textrm{em}}(k^2)}{f_\rho m_\pi} A_0(m_\pi^2)  \ ,
  \end{equation}
  where $f_\rho = 0.209$ GeV and $A_0(m_\pi^2) \simeq A_0(0)= 0.36\pm 0.04$\cite{ Bal04, Str15}.

A similar procedure can be applied to the $I = 0$ channel, assuming dominance of a scalar resonance,  $B \to S_0 \to \pi \pi$. We  
write
\begin{align}
\left\langle \pi^-(k_1)\pi^+(k_2)|\bar{u} \gamma_\nu(1-\gamma_5) b|B^-\right\rangle{}& =\left\langle \pi^-(k_1)\pi^+(k_2) |S^0\right\rangle \mathcal{B}_S(k) \left\langle S^0(k) |\bar{u}\gamma_\nu(1-\gamma_5)b|B^- \right\rangle \ ,\nonumber 
\end{align}
where the relevant part of the form factor 
for the $B \to S_0$ transition is defined as (see e.g. \cite{Che05})
\begin{equation}
\left\langle S^0(k) |\bar{u}\gamma_\nu(1-\gamma_5) b|B^-(p_B) \right\rangle= -i q_\nu \frac{m_B^2-m_S^2}{q^2} F_0^{B S} (q^2)  \ ,
\end{equation}
{with}  $q = p_B-k$.
Similarly, we can write $F_\pi^S$ in the same way
\begin{equation} \label{eq:Fpismod}
F_\pi^S (k^2)=  \frac{(m_u+m_d)}{m_\pi^2} f_S m_S^2 g_{S\pi^-\pi^+} \mathcal{B}_S(k^2) \ ,
\end{equation}
where the decay constant and the strong coupling constant are defined by
	\begin{equation}
	\left\langle S^0|\bar{u}u + \bar{d}d|0\right\rangle = f_S m_S \ , \; \; 	\left\langle \pi^-(k_1)\pi^+(k_2) |S^0\right\rangle = g_{S\pi^-\pi^+} m_S \ .
	\end{equation}
Finally, we substitute again $g_{S\pi^-\pi^+}\mathcal{B}_S$ for $F_\pi^S$. However, $F_0^{BS}$ and $f_S$ are unknown for the lightest scalar resonances. {W}e thus model the isoscalar form factor through
\begin{equation}\label{eq:Ft0mod}
F_t^{I=0}(k^2, k_3\cdot\bar{k}) = \frac{m_B^2}{m_\pi f_\pi} \beta e^{i\phi} F_\pi^S (k^2) \ ,
\end{equation}
where the model parameters $\beta$ and $\phi$ can be obtained from {a fit to} data. The approximations made to obtain the $B\to \pi\pi$ form factors in terms of the pion form factors are currently unavoidable. In the future this modelling might be circumvented using QCD sum rules that employ the pion distribution amplitudes~\cite{Ham15,Che172}. {In addition, these models can be fitted separately to the light-cone sum rules
with $B$ distribution amplitudes, as done in \Ref{Che17}.}

\section{CP violation in  $B^-  \to \pi^-  \pi^+ \pi^-$} 
\label{sec:CPV}

\begin{figure}
  \centering
\subfloat[Dalitz distribution of the LHCb data]{\includegraphics[width=0.47\textwidth,height=6.5cm]{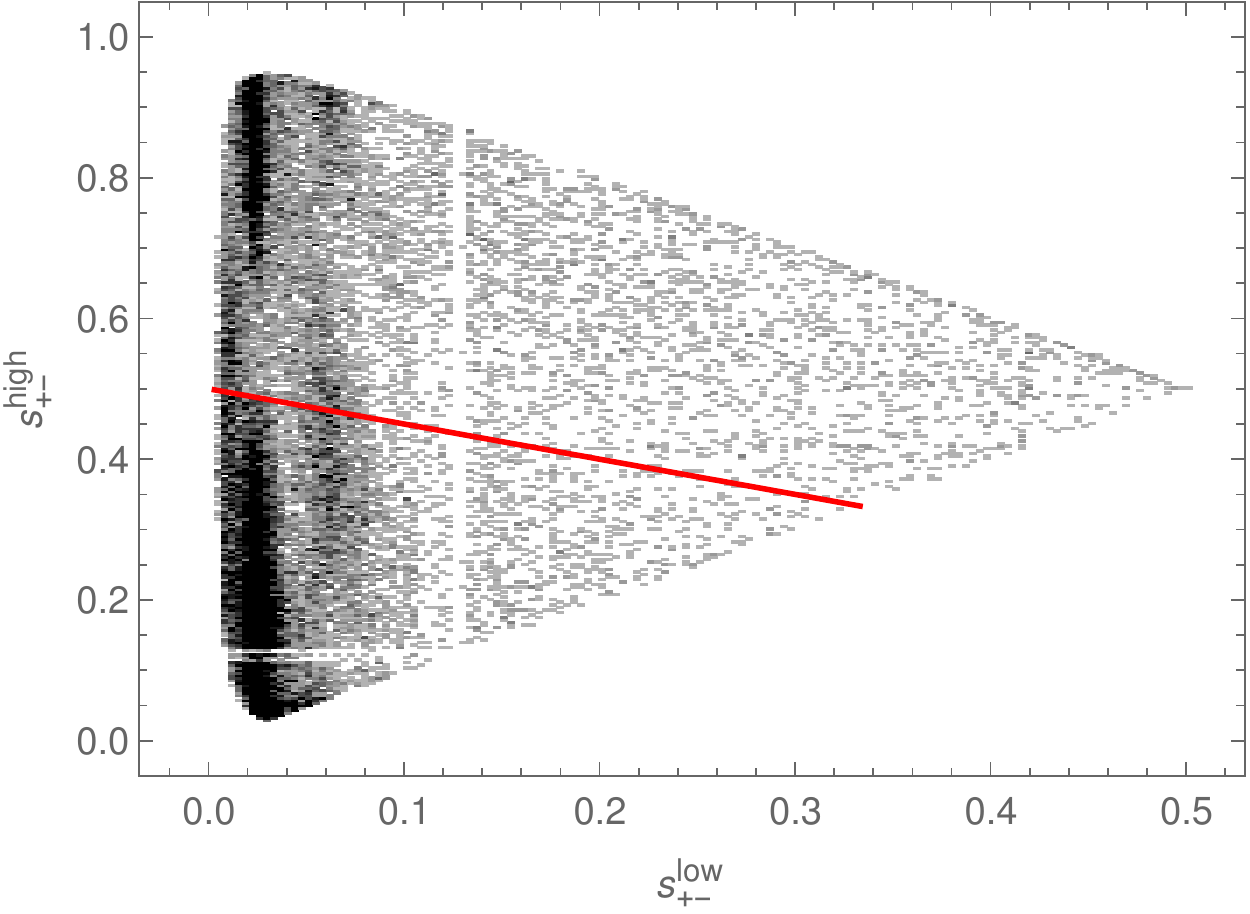}\hspace{4mm}}
   \label{fig:charmmoddal}\subfloat[Dalitz distribution of our model]{
\includegraphics[width=0.47\textwidth,height=6.5cm]{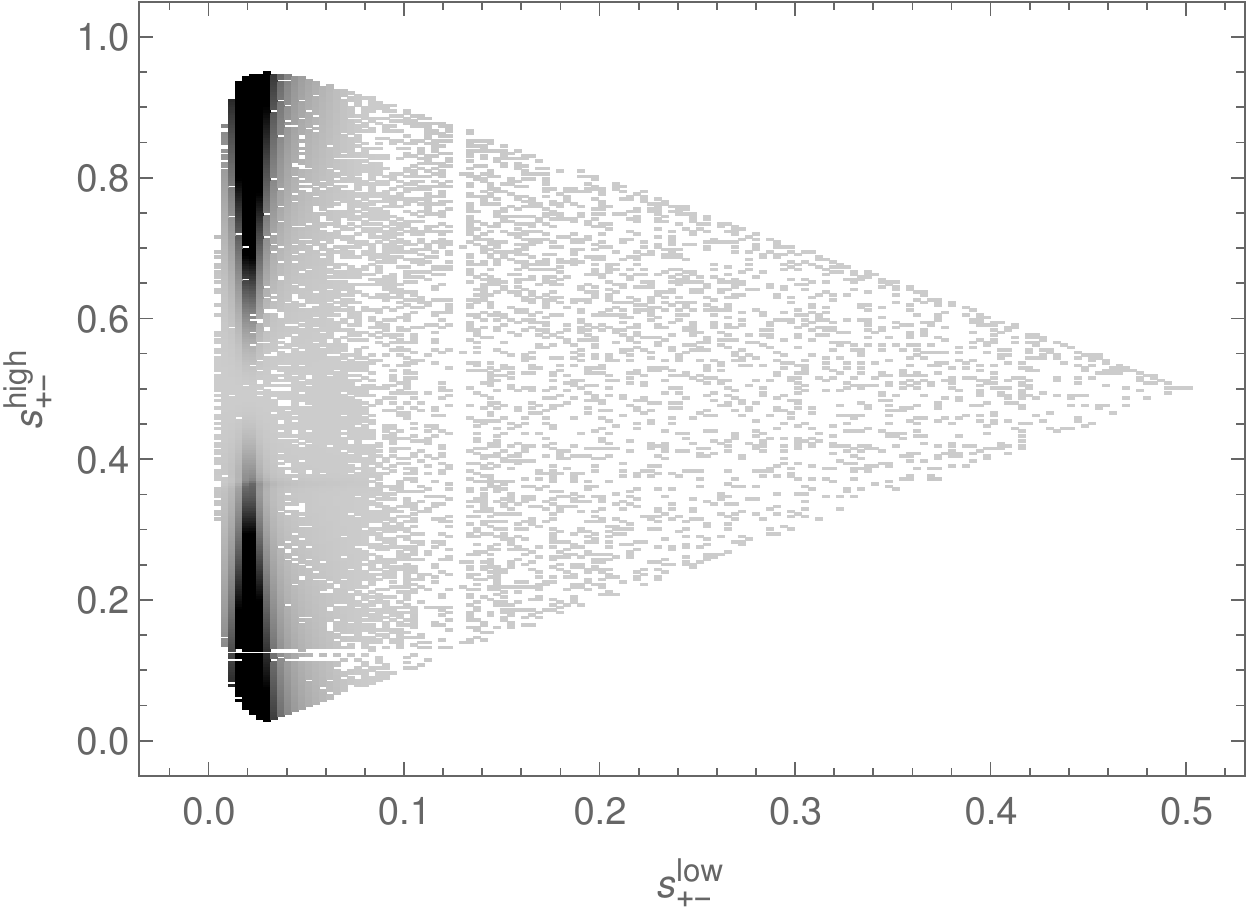}}
 \caption{Dalitz distribution for $B^- \to \pi^-\pi^+\pi^-$ (a) as measured by the LHCb Collaboration \cite{LHCbdata} where the region below the line corresponds to $\cos\theta_\pi>0$ line as discussed in the text (b) Dalitz distribution of our model including (\ref{mod1}).}
  \label{fig:LHCb}
\end{figure} 

We start from the amplitude in Eq.~\eqref{eq:ampli}, use $F_t^{I=1}$ in (\ref{eq:Ft1mod}) and we model $F_t^{I=0}$ using (\ref{eq:Ft0mod}).
%
%
Our model thus {contains only two free parameters:} $\beta$ and the phase $\phi$. 

The CP asymmetry is {given by
\begin{equation}\label{eq:CPas}
A_{\rm CP}(s_{\pm}^{\rm low} , s_{\pm}^{\rm high}) =  \frac{|\mathcal{A}(s_{\pm}^{\rm low} , s_{\pm}^{\rm high})|^2 - |\bar{\mathcal{A}}(s_{\pm}^{\rm low} , s_{\pm}^{\rm high})|^2}{|\mathcal{A}(s_{\pm}^{\rm low} , s_{\pm}^{\rm high})|^2 + |\bar{\mathcal{A}}(s_{\pm}^{\rm low} , s_{\pm}^{\rm high})|^2}  \ ,
 \end{equation}
where $\bar{\mathcal{A}}$ is equal to $\mathcal{A}$ with all weak phases conjugated.}
The required weak phase difference is given through the different structures with $\lambda_u = V_{ub} V_{ud}^* = |V_{ud} V_{ub}| e^{-i \gamma}$, where $\gamma$ is the corresponding weak phase of the Unitarity Triangle and
$\lambda_c = V_{cb} V_{cd} = |V_{cb} V_{cd}|$ is real {within our convention}. We {will} use the values {quoted} in \cite{Kop17}.

At tree level $a_4^u=a_4^c$ and the coefficients $a_i$ are given by the Wilson coefficients $C_i$ \cite{BBNS, Ben01}
\begin{equation}
a_{1,2,4} = C_{1,2,4} + \frac{C_{2,1,3}}{N_C} \ , 
\end{equation}
where $N_C= 3$ denotes the number of colors. At $\mathcal{O}(\alpha_s)$ the coefficients $a_i$ also {acquire} perturbative strong phases \cite{BBNS, Ben09, Bel15}. These can be included using the partial QCD factorization formalism {discussed} in \Ref{Kra15}, which requires taking into account the convolutions of the hard kernels with the generalized $2\pi$ distribution amplitude (DA) (see also \cite{Poly98, Die00, Die03}). At leading order, the pion DA  and the generalized $2\pi$ DA reduce to their local limits, corresponding to {Eqs.}~(\ref{Pidec}) and (\ref{eq:2pionconst}), respectively. Since these $\mathcal{O}(\alpha)$ correction cannot generate large CP asymmetries, we work at {leading order, where the coefficients $a_i$ are real}, leaving higher-order effects for future studies. The required strong phase difference to generate CP violation should thus come from the interference between the form factors $F_\pi^{\textrm{em}}$ and $F_t$. 

In our model, the phase{s} of $F_\pi^{\textrm{em}}$ and $F_t^{I=1}$ are identical. For elastic scattering {(below the threshold of the first inelasticity in $\pi\pi$ scattering)} this is a general statement following from Watson's theorem. {This condition has been emphasized within the framework of QCD sum rules in \Ref{Che17}.} 

We define, as before, the strong phases $\delta_S$ and $\delta$ as
\begin{equation}
F_\pi^S = |F_\pi^S(s_{\pm}^{\rm low})| e^{i\delta_S (s_{\pm}^{\rm low})} \ , \;\;\;\; F_\pi^{\rm em} = |F_\pi^{\rm em}(s_{\pm}^{\rm low})| e^{i\delta(s_{\pm}^{\rm low})} \nonumber \ .
\end{equation}
{Inserting the amplitude in Eq.~(\ref{eq:ampli}) into (\ref{eq:CPas}), one finds that the} CP asymmetry is proportional to
 \begin{equation}\label{eq:acpeasy}
 A_{CP}(s_{\pm}^{\rm low}, \cos\theta_\pi)= \beta \,\sin\gamma \,\sin(\delta_S(s_{\pm}^{\rm low})+\phi-\delta(s_{\pm}^{\rm low})) \,\cos\theta_\pi\,  |F_\pi^S(s_{\pm}^{\rm low})|\, |F_\pi^{\rm em}(s_{\pm}^{\rm low})|\, g(s_{\pm}^{\rm low}) \ ,
 \end{equation}
where $g(s_{\pm}^{\rm low})$ is a real function that can be computed from Eqs.~(\ref{eq:ampli}) and (\ref{eq:CPas}). We have replaced the $s_{\pm}^{\rm high}$ variable with $\cos\theta_\pi$ following {Eq.~}(\ref{eq:cos}). We see that only the interference between $F_\pi^{\textrm{em}}$  and $F_t^{I=0}$ terms contribute to the CP asymmetry. Therefore, the specific parametrizations for  $F_\pi^{\textrm{em}}$ and $F_t^{I=0}$ are of crucial importance. Here we use the parametrizations {discussed in Section~\ref{sec:NPinput}} and depicted in Figs.~\ref{fig:vectorff} and~\ref{fig:ff}, which allow us to perform a first analysis of our QCD-based model. The model dependence of our approach could be reduced in the future when more data for the form factors is available. We do not take into account uncertainties for the pion form factors.  

\begin{figure}
  \centering
  \subfloat[]{
\includegraphics[width=0.49\textwidth,height=6cm]{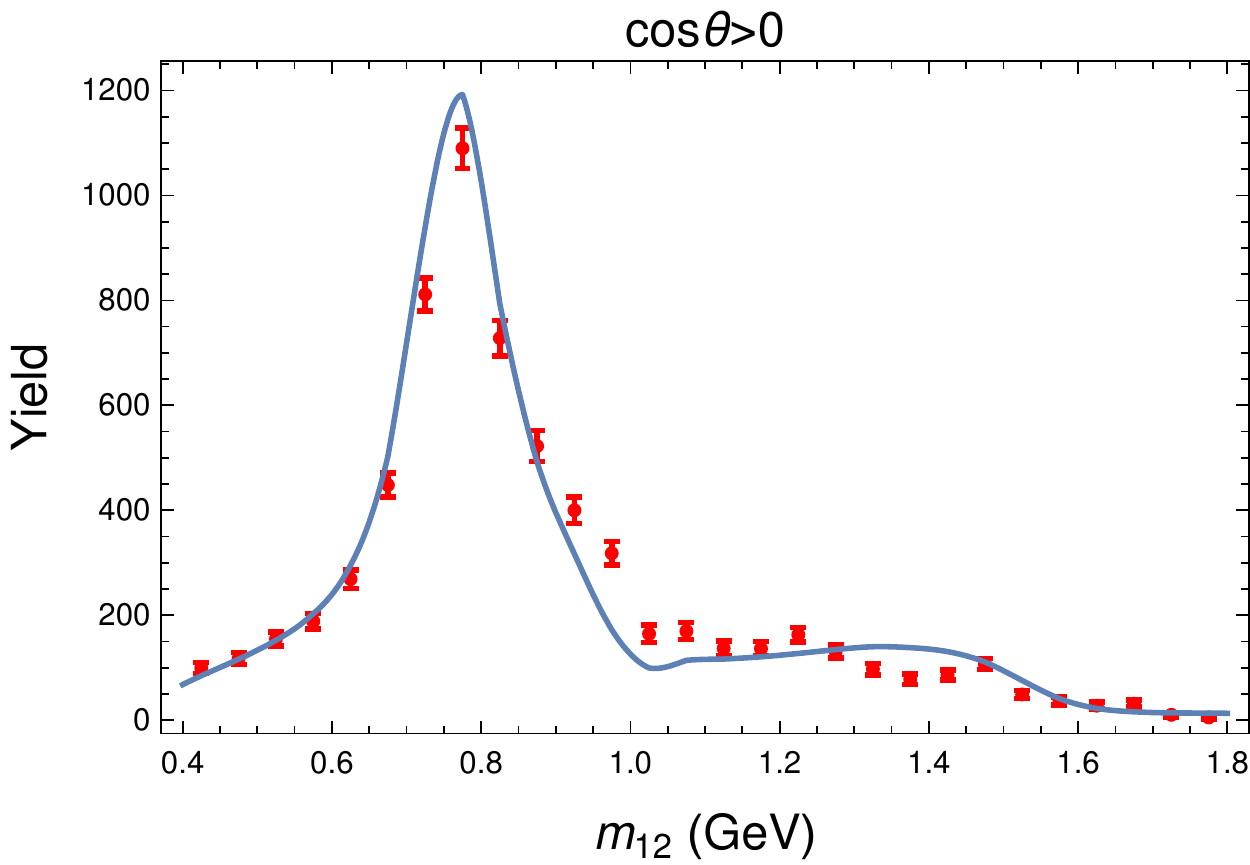}\hspace{3mm}}
    \subfloat[]{\includegraphics[width=0.48\textwidth,height=6cm]{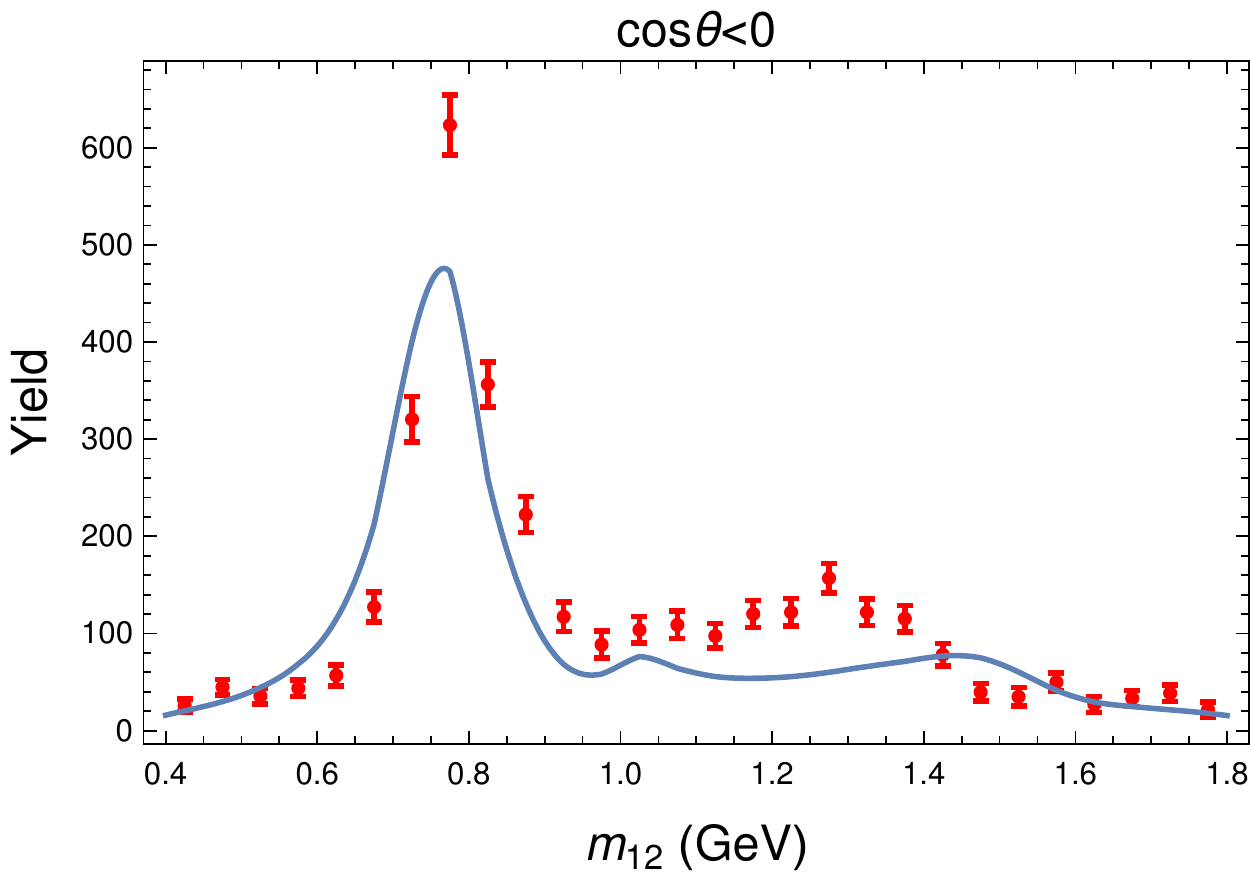}} \\
      \subfloat[]{
\includegraphics[width=0.485\textwidth,height=6cm]{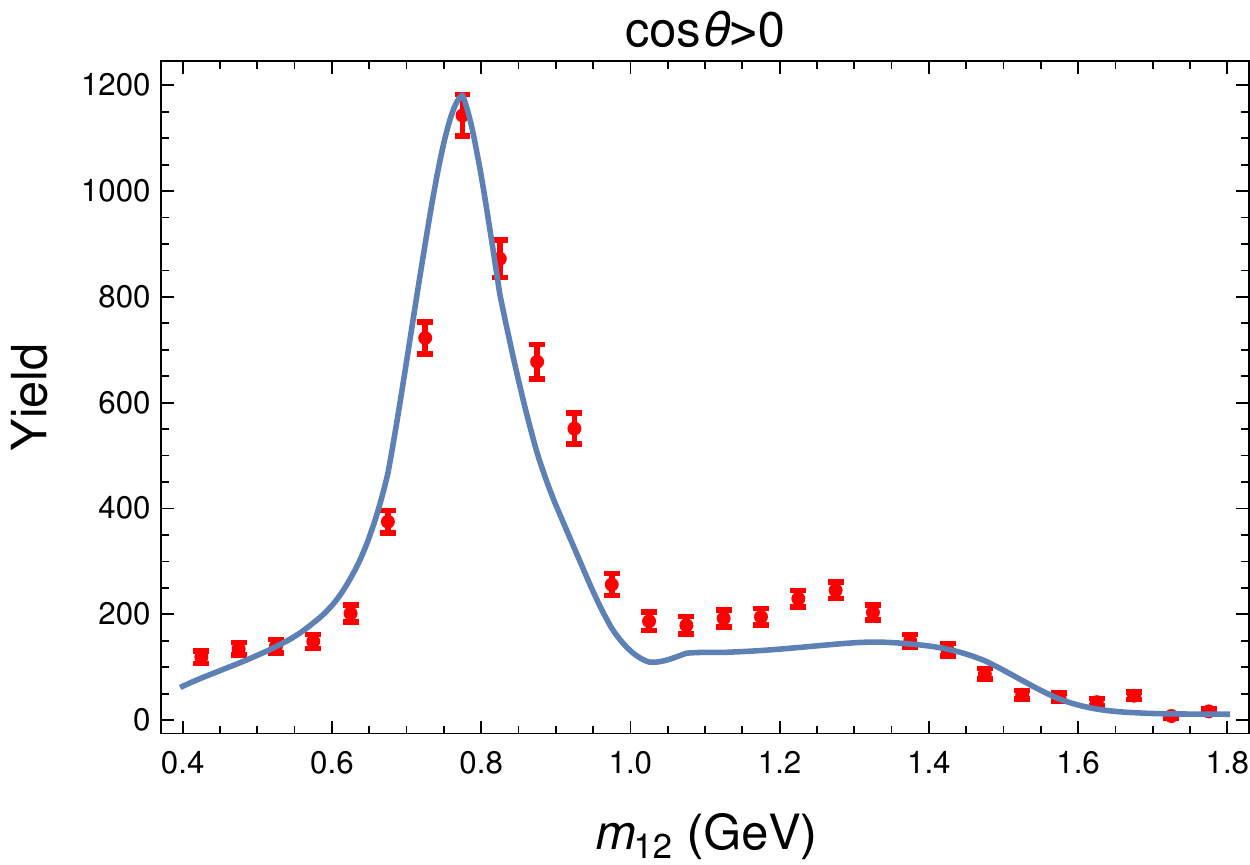}\hspace{3mm}}
    \subfloat[]{\includegraphics[width=0.48\textwidth,height=6cm]{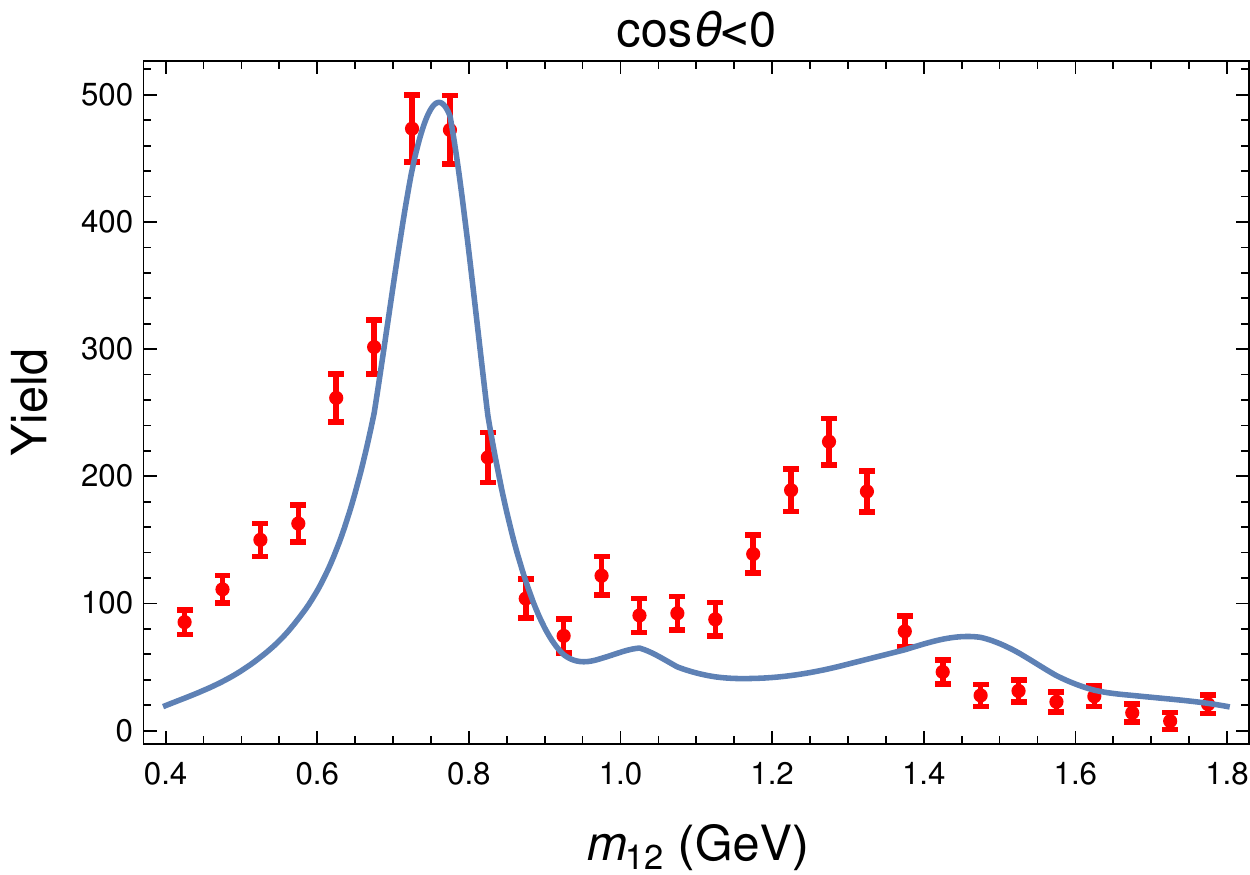}} 
  \caption{Best fit of our model for (a, b) $B^+$ and (c, d) $B^-$ and the LHCb projections~\cite{LHCbdata} as a function of $m_{12}$.}
  \label{fig:fits}
\end{figure}

Using the data from the LHCb Collaboration~\cite{LHCbdata}, we may fit our model parameters $\beta$ and $\phi$ {directly}.
Unfortunately, the {full} efficiency- and background-corrected Dalitz distribution {is not provided by the LHCb analysis. Therefore, we use} the projections of the data given for $B^+$ and $B^-$ decays, separated for $\cos\theta_\pi<0$ and $\cos\theta_\pi>0$. We show these two regions in the $B^-\to \pi^-\pi^+\pi^-$ Dalitz distribution {in Fig.~\ref{fig:LHCb}}, as given in \Ref{LHCbdata}, where $\cos\theta_\pi>0$ corresponds to $s_{\pm}^{\rm high}<\tfrac{1}{2}(1-s_{\pm}^{\rm low})$, i.e. the lower part of the distribution. Fig.~\ref{fig:fits} shows the projections of the LHCb data for $\cos\theta_\pi<0$ and $\cos\theta_\pi>0$ in bins of $0.05$ GeV {for} the variable $m_{12}= k_1+k_2$.

{We now perform a fit to the data to determine the most likely values for the model parameters.
The fit is performed by a standard $\chi^2$ minimization.} Our model predicts the decay rate for each bin; since the measurement of the absolute branching ratio is not available we have to scale our results to match the arbitrary units used in Fig.~\ref{fig:fits}. Fitting this scaling parameter together with our parameters $\beta$ and $\phi$ gives: 
\begin{equation}
\beta= 0.18 \;\;\; \textrm{ and} \;\;\; \phi=18^\circ \ . 
\end{equation}
The yield predictions with these best fit parameters are also {included} in Fig.~\ref{fig:fits}. These figures show that our fit represents the data for $B^+$ at $\cos\theta>0$ best, although in general our fit describes the data very poorly. We therefore refrain from giving an error {to} our fit parameters.
These results call for refinements in the modelling of the form factors, which at this stage has been relatively simplistic. A number of possibilities will be mentioned later.

\begin{figure}
 \centering
 \subfloat[]{
\includegraphics[width=0.48\textwidth,height=6.5cm]{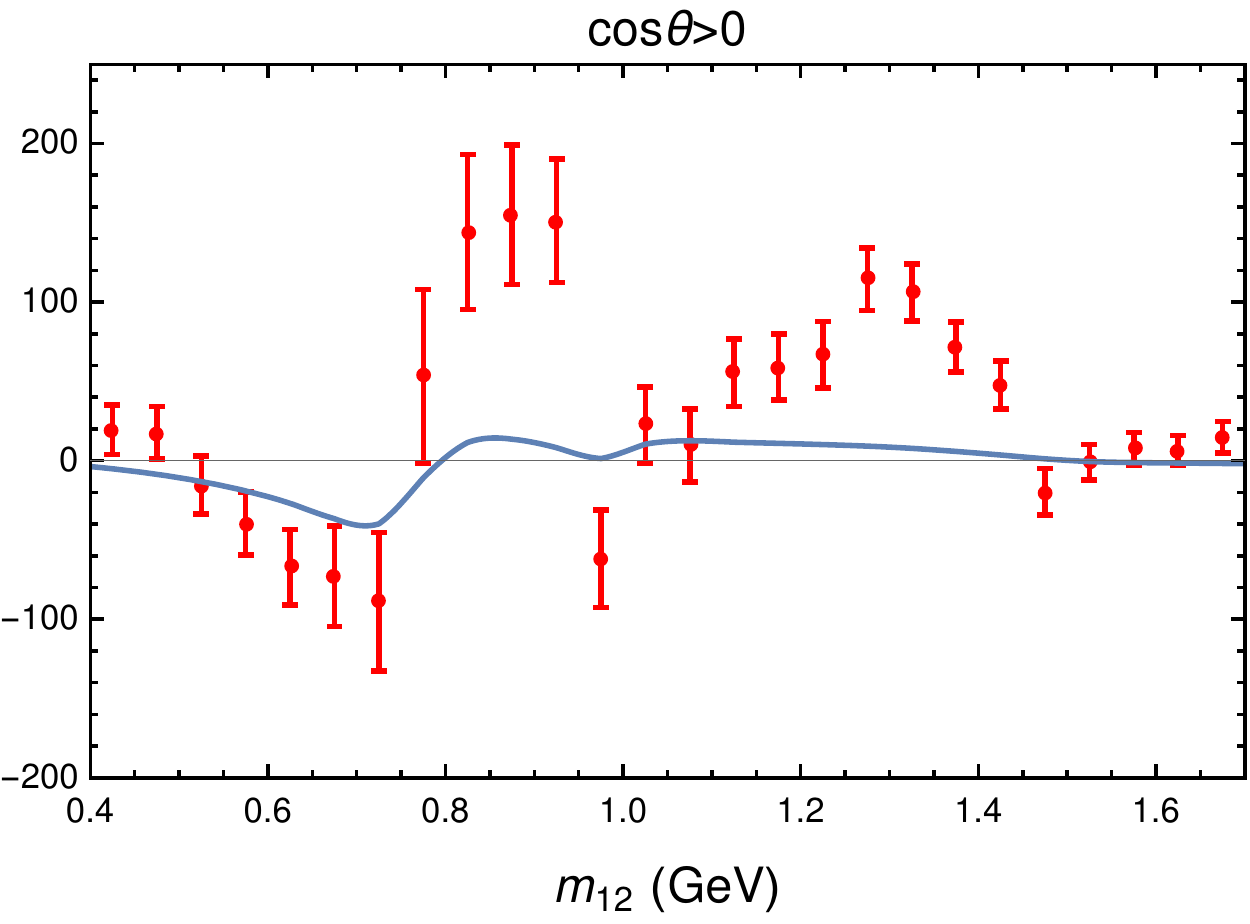}\hspace{3mm}}
  \subfloat[]{
\includegraphics[width=0.48\textwidth,height=6.5cm]{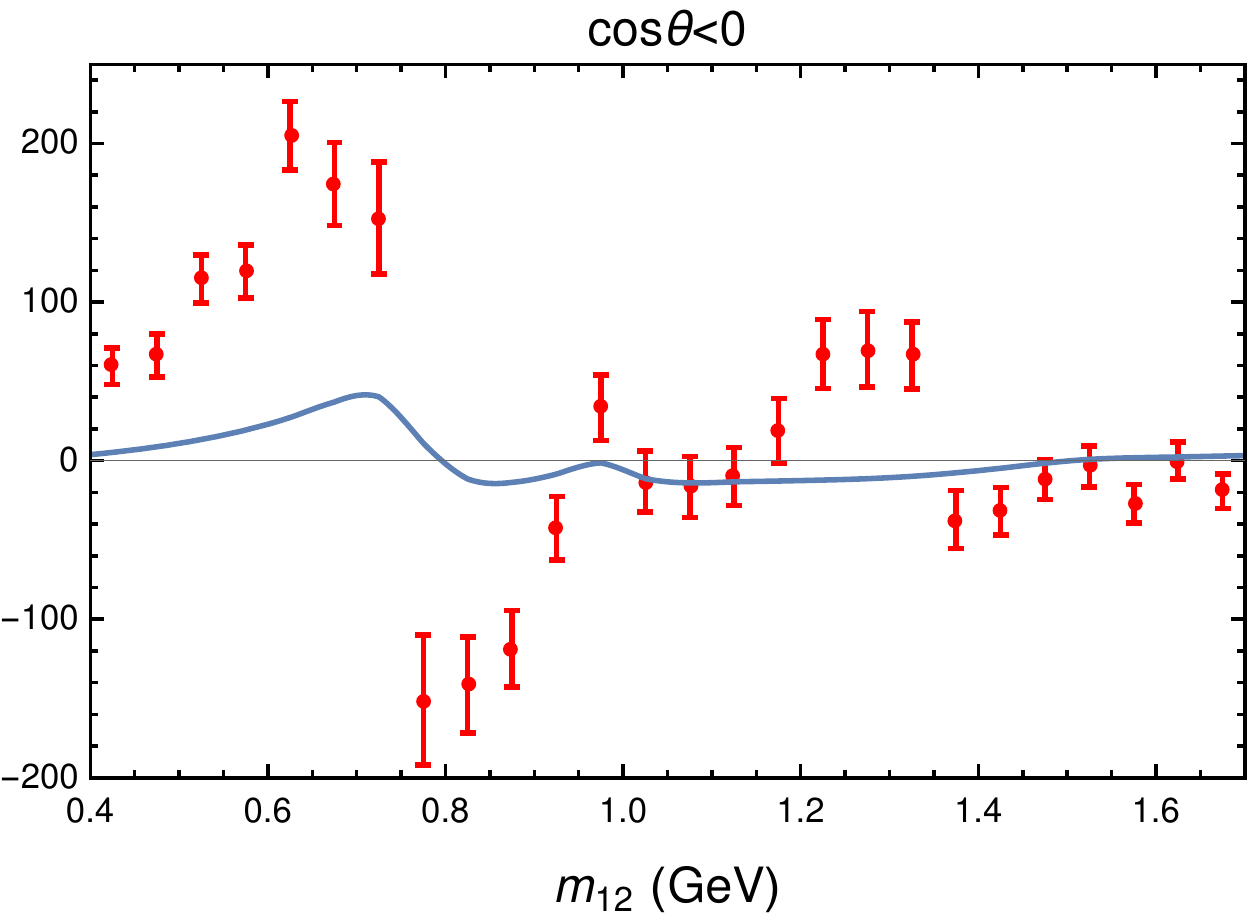}}
 \caption{Difference between the $B^-$ and $B^+$ yield in our best fit compared to the LHCb data \cite{LHCbdata}.}
 \label{fig:BminBplus}
\end{figure}

{A more clear picture of the situation is obtained by scrutinizing the CP asymmetry in more detail.}
In Fig.~\ref{fig:charmmodcp} {we show the complete CP distribution as provided by LHCb~\cite{LHCbdata}} in a specific binning that ensures that each bin has the same number of events. {The} projections for the $B^-$ - $B^+$ {yield differences} are also given by LHCb~\cite{LHCbdata}. We {show these} in Fig.~\ref{fig:BminBplus} {together} with the outcome of our {fit}. The {resulting} CP violation in our model is much smaller than that seen in the data, nevertheless it reproduces the gross structures except for the region around $1.3$ GeV. 

In the region around $m_\rho$ we expect our model to most accurate. The differences as seen in the CP asymmetries might be due to the simplistic model used for $F_t^{I=1}$ in (\ref{eq:Ft1mod}). To study the effect of relaxing this assumption, we added another fit parameter to $F_t^{I=I}$. Performing then the $\chi^2$ analysis, leads to a slightly better agreement around the $\rho$ peak, but the total fit still remains a poor description of the data. The neglected higher-order terms might also give small modifications in this region. However, our model clearly fails to describe the interesting behavior of the CP asymmetry around $1.3$ GeV. Here there is a positive CP asymmetry for both of the $\cos\theta$ regions. In our model, the small CP violation in this region switches sign as does the CP asymmetry in the $\rho$ region. This is because $A_{CP}$ in {Eq.~}(\ref{eq:acpeasy}) is only generated by a vector-scalar interference, {which} always comes with a $\cos\theta_\pi$ term, and hence the CP asymmetry always switches sign when comparing $\cos\theta>0$ and $\cos\theta<0$ (see also \cite{Bed09, Nog15, Nog16} for an elaborate discussion on these {issues}). However, if the CP asymmetry {were} dominated by two $S$ or $S$-$D$ wave interferences, the CP asymmetry would not switch sign, which could be an explanation for the behaviour in this region. Additional $S$ or $D$-wave terms might still arise in our approach when including higher-order (twist) corrections, we leave the study of these corrections for future work. In addition, we note that this region is also at the boundary of where the scalar form factor depicted in Fig.~\ref{fig:ff} can be trusted. {Therefore, inelasticities may also play an important role.}

For this first study, we have compared to the available projections of the LHCb data. Therefore, important information about the CP asymmetry in the $s_{+-}^{\rm high}$ variable is essentially washed out. Our simple model does not give a good quantitative description of the CP asymmetries, however, several refinements are possible. For {future studies it would be beneficial and desirable to have} the full information on the Dalitz and CP distributions.

\section{{Conclusion} and comment on charm resonances}

\begin{figure}
 \centering
 \label{fig:cplhcb}\subfloat[CP Distribution from LHCb \cite{LHCbdata}]{
\includegraphics[width=0.465\textwidth,height=6.45cm]{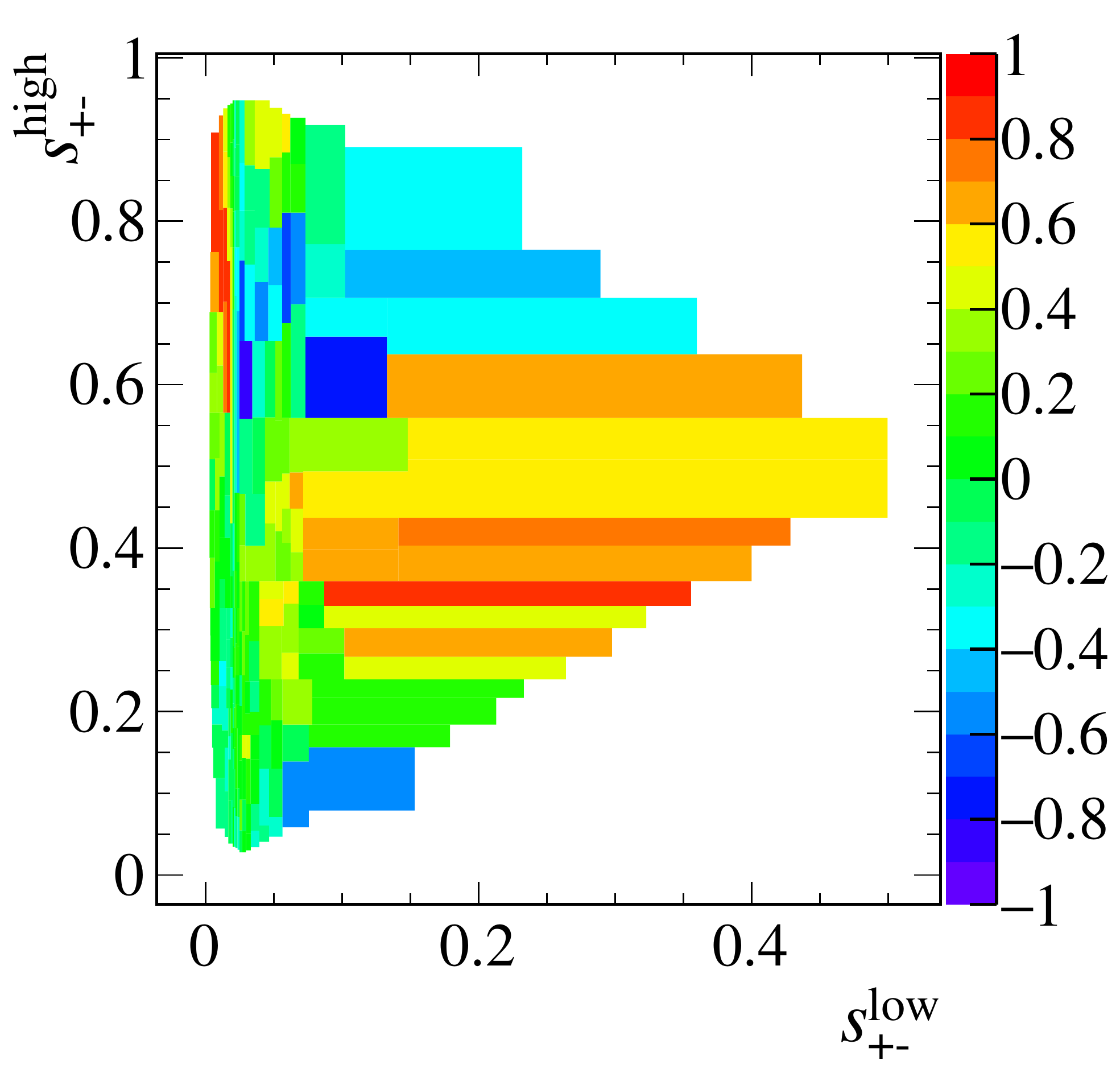}\hspace{2mm}}
\subfloat[CP distribution for our model]{\raisebox{0.2cm}{
\includegraphics[width=0.465\textwidth,height=5.95cm]{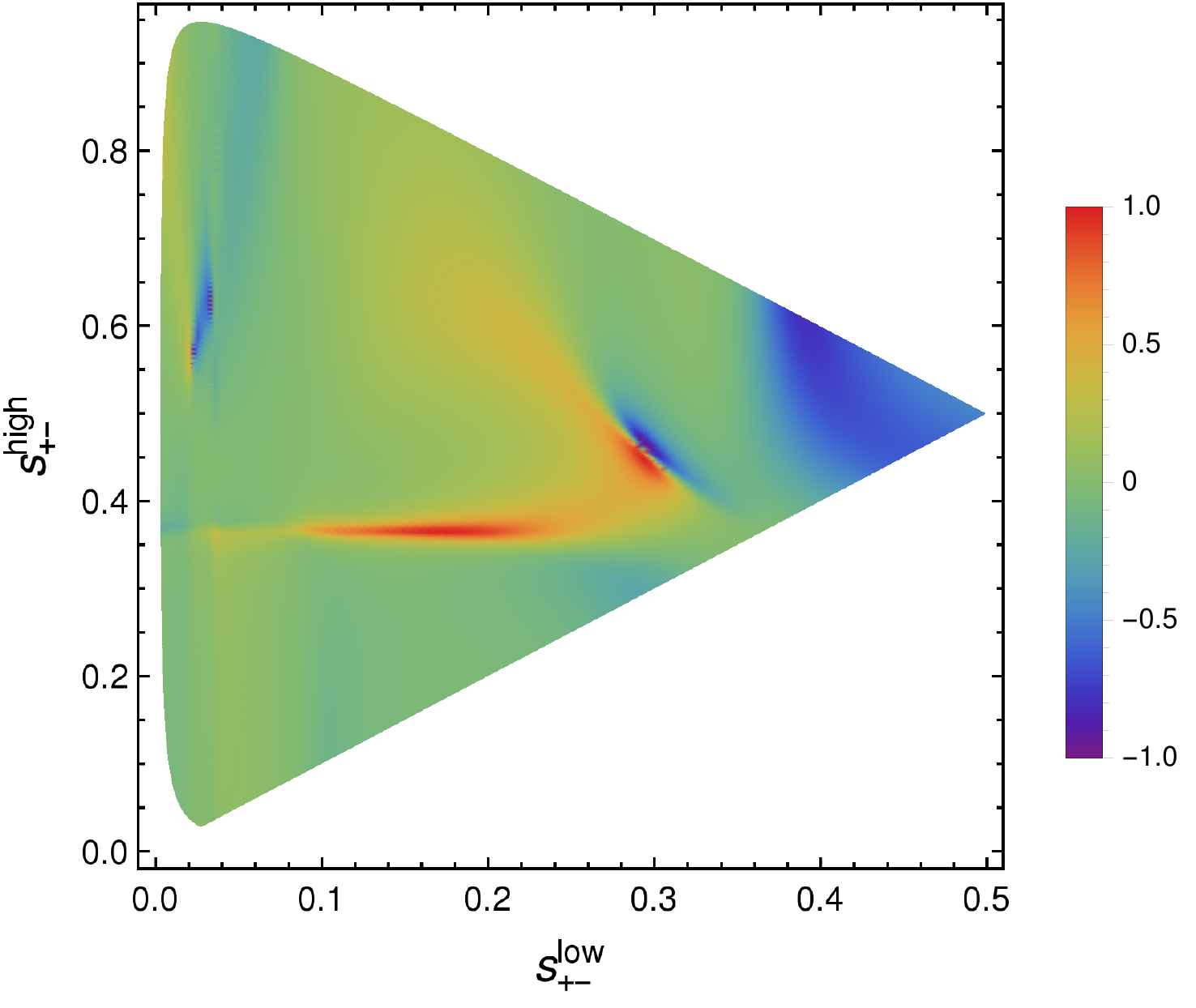}}}
 \caption{CP distribution for (a) the LHCb data~\cite{LHCbdata} and (b) our model including the charm-resonance structure of (\ref{mod1}).}
 \label{fig:charmmodcp}
\end{figure} 

We {have} discussed a data-driven model based on QCD factorization to study CP violation in $B \to \pi\pi\pi$, which depends on the model parameters $\beta$ and $\phi$ {(a strong phase)}.
The form factors for the $B\to \pi \pi$ transition as well as the time-like pion form factors have non-perturbative strong phases that lead to a complicated phase structure of the 
amplitudes $T_u$ and $T_c$. Although we {have shown} that our simple model can describe some of the features of the decay rates and CP asymmetries, it cannot capture all the physics which is relevant 
for the local CP asymmetries. We have discussed some possible refinements of our model to accommodate these features.
Nonetheless, {beyond the particular model-dependent choices adopted in this analysis,
the aim of being able to fix the amplitude in Eq.~(\ref{eq:ampli}) completely from data on the
time-like pion form factors is an important one. In this way one can avoid the use of}
isobar assumptions \cite{Isobar1, Isobar2} and Breit-Wigner-shaped resonance models (e.g. \cite{Che051, Che07,Che16, Wan15}). {We are confident that progress will be made in this
direction.}

\bigskip

Since we use only the parametrizations
of the scalar and vector form factors of the pion, the {modelled} non-perturbative phases
are only the ones {related to} the final-state interactions of the two {opposite}-sign pions.
This means that we can only expect this simple model to work within the regions where this is the dominant effect, i.e. at the corresponding edges of the phase space.
Nevertheless, one might consider a possible extension of our model, especially when considering the {measured} CP asymmetry in Fig.~\ref{fig:charmmodcp}.
These measurements find large local CP asymmetries at high $s_{+-}^{\rm low}$ and in regions of the phase space where there seem to be not many events when comparing with the Dalitz distribution in Fig.~\ref{fig:LHCb}. Unfortunately, projections of this high momentum region are not (yet) available. 

It is possible to extrapolate the pion form factors up to larger invariant mass{. H}owever, since there is no extra {``structure'' in this region} this would {fix} the phases {to around} $180^\circ$ everywhere, {suppressing the local CP asymmetry} at high $s_{+-}^{\rm low}$ in contrast to the observation. The observed CP asymmetry might be created by subleading effects that were thus far assumed to be suppressed, but that might give significant effects at such high momenta.
We note that the amplitudes $T_u$ and $T_c$ differ by the fact that $T_c$ contains 
{penguins with charm, and it is thus sensitive to} the heavy charm-quark mass.
Subleading terms in QCD factorization for two-body decays generate perturbatively calculable strong phases for the coefficient $a_4$ which generates the CP violation in the $B\to\rho\pi$ decay. 
In the three-{body} decay one might expect a similar effect from the charm quarks, which would 
modify the shape of the local CP asymmetry. The details of this contribution will depend on the non-perturbative interaction of the two charm quarks in~$T_c$.

Clearly we do not have a way to actually compute this, so we have to make use of some modelling 
to get a qualitative picture. {We first regard the region close to the charm threshold ($2 m_c$)
as the relevant region where sharp charmed resonances may affect the CP asymmetry.}
The simplest way to introduce non-trivial phases is to consider a  resonance-like structure in $T_c$ described by a Breit-Wigner shape. Thus we modify our model {by a simple addition:}
\begin{eqnarray}
T_c &=& T_c^{(0)}  + g \frac{4 m_c^2}{m_{+-}^2 - 4 m_c^2 + i m_c \Gamma}   \ ,
\label{mod1}\\
T_u  &=& T_u^{(0)}  \ ,  
\end{eqnarray}  
where $T^{(0)}_q$ is the leading order amplitude given in Eq.~(\ref{eq:ampli}) {by the term proportional to $\lambda_q$}. As an example, we fix the constant $g$ to be  $0.02$ and $\Gamma = 0.15$ GeV, {and take} $m_c = 1.6$ GeV {for definiteness}. 

In Fig.~\ref{fig:LHCb} we show the resulting logarithmic Dalitz distribution, {compared to the measured} Dalitz distribution. Clearly the Dalitz plot is dominated by the 
resonance structure {given by} the time-like pion form factor, {and} 
the subleading term modelled by Eq.~(\ref{mod1}) yields indeed small contributions as expected (and can be tuned with the constant $g$). 
However, {as mentioned earlier it is not possible to} qualitatively compare our Dalitz distribution with the {measured one}~\cite{LHCbdata} because the latter is not background subtracted. 
{A more thorough comparison in the line of that in Section~\ref{sec:CPV} would require the} data projections for the high momentum part as well. 

In Fig.~\ref{fig:charmmodcp} we also show the corresponding local CP asymmetry distribution. The subleading term (\ref{mod1}) now generates a non-perturbative, phase-space dependent phase difference between $T_c$ and $T_u$ which induce{s} sizeable local CP asymmetries in the region {around} the charm threshold. We note that the actual CP distribution depends on the values of $g, \Gamma, \beta$ and $\phi$. It might be interesting to include this charm-resonance model into an amplitude analysis to obtain further insights on the behaviour {on} the CP asymmetry at high momenta. 

Obviously this is only a crude model{. H}owever, we note that the qualitative structure is in agreement with the observations by LHCb~\cite{LHCbdata}. The data are not yet very precise, but the sizeable CP asymmetries  observed by LHCb are compatible with structures originating from charm-threshold effects as we model them in {Eq.~}(\ref{mod1}). However, it {is extremely} difficult to achieve a quantitative understanding of these effects from QCD. {The issue of the charm contributions has also been a hot topic of discussion in two-body decays, but it is in three-body decays where there might be a chance to measure their effect and to interpret it. We believe this qualitative discussion may provide some motivation.}

\section*{Acknowledgements}

This work has been funded by the Deutsche Forschungsgemeinschaft (DFG) within research unit FOR 1873 (QFET). K.V. would like to thank the Albert Einstein Center for Fundamental Physics (AEC) in Bern for its hospitality and support. {J.V. acknowledges funding from the Swiss National Science Foundation, and from the European Union's Horizon 2020
research and innovation programme under the Marie Sklodowska-Curie grant
agreement No 700525 `NIOBE'.}
\appendix

\end{document}